\documentclass[12pt]{article}
\usepackage{amsmath,amsfonts,amssymb,graphicx}
\usepackage{graphics, setspace}
\usepackage{cite} 
\usepackage{hyperref}
\usepackage{epstopdf}
\DeclareGraphicsExtensions{.eps}

\begin{document}
\thispagestyle{empty}

\def\theequation{\arabic{section}.\arabic{equation}}
\def\a{\alpha}
\def\b{\beta}
\def\g{\gamma}
\def\d{\delta}
\def\dd{\rm d}
\def\e{\epsilon}
\def\ve{\varepsilon}
\def\z{\zeta}
\def\B{\mbox{\bf B}}\def\cp{\mathbb {CP}^3}

\newcommand{\h}{\hspace{0.5cm}}

\begin{titlepage}

\renewcommand{\thefootnote}{\fnsymbol{footnote}}
\begin{center}
{\Large \bf Currents algebra for an atom-molecule Bose-Einstein condensate model}
\end{center}
\vskip 1.2cm \centerline{\bf Gilberto N. Santos Filho$^{1}$}

\vskip 10mm
\centerline{\sl$\ ^1$ Centro Brasileiro de Pesquisas F\'{\i}sicas - CBPF}
\centerline{\sl Rua Dr. Xavier Sigaud, 150, Urca, Rio de Janeiro - RJ - Brazil}
\vskip .5cm

\centerline{\tt gfilho@cbpf.br}
 
\vskip 20mm

\baselineskip 18pt

\begin{center}
{\bf Abstract}
\end{center}

I present an interconversion currents algebra for an atom-molecule Bose-Einstein condensate model and use it to get the quantum dynamics of the currents. For different choices of the Hamiltonian parameters I get different currents dynamics. 

\end{titlepage}
\newpage
\baselineskip 18pt

\def\nn{\nonumber}
\def\tr{{\rm tr}\,}
\def\p{\partial}
\newcommand{\non}{\nonumber}
\newcommand{\bea}{\begin{eqnarray}}
\newcommand{\eea}{\end{eqnarray}}
\newcommand{\bde}{{\bf e}}
\renewcommand{\thefootnote}{\fnsymbol{footnote}}
\newcommand{\be}{\begin{eqnarray}}
\newcommand{\ee}{\end{eqnarray}}

\vskip 0cm

\renewcommand{\thefootnote}{\arabic{footnote}}
\setcounter{footnote}{0}

\setcounter{equation}{0}
\section{Introduction}

 Since the first experimental verification of the Bose-Einstein condensation (BEC) \cite{ak,anderson,wwcch}, occurred more then seven decades after its theoretical prediction \cite{bose,eins},  a great effort in the theoretical and experimental viewpoint has been made in the study of this quantum many body physical phenomenon \cite{Hulet,Dalfovo,l01,Bagnato,cw,Piza,Bloch,Carusotto}. Looking for new applications an atom-molecule Bose-Einstein condensates was experimentally produced applying magnetic field near a Feshbach resonance in an atomic BEC \cite{Donley,Zoller}.  The magnetic field pulses produces a coherent interconversion dynamics of atoms and molecules. Another techniques used to produce an atom-molecule Bose-Einstein condensate are the photoassociation \cite{Souza} and the two-photons Raman transition \cite{Wynar}.  These experimental opened the possibility of the cold and ultracold quantum chemistry  \cite{Wynar,Richter,Hutson,Regal,Ospelkaus,Heinzen,Goral,Carr,Kevin,Makrides,Dulieu,Pillet,Amelink,Pillet2,Durr,GSantos09 }.  I am considering here an atom-molecule Bose-Einstein condensate model, used to study this coherent interconversion dynamics of atoms and homonuclear diatomic molecules \cite{GSantos06a,GSantos10}.  This model is integrable in the sense that it can be solved by the quantum inverse scattering method (QISM) \cite{GSantos11a,GSantos11b,jlletter,jlreview,jinter,ATonel,jlsigma,Angela2,Angela,GSantos06b,GSantosBVS-PLB,jdensity,jbroots,GSantos13}. In this work I will discuss the symmetries and the  interconversion currents algebra of the model and use they to study the quantum dynamics of the currents.  This method was applied in the study of the Josephson tunnelling phenomenon in a two-site Bose-Hubbard model to get the quantum dynamics of the tunnelling currents of atoms as a function of the parameters of the Hamiltonian \cite{GSantosCA1,GSantosCA2}. We can also apply this method to a model recently used to study the   tunnelling between two atom-molecule BECs coupled by Josephson tunnelling \cite{Motohashi1,Motohashi2,Motohashi3,Motohashi4,Dukelsky}. It is also worth to note that we can considerer the atom-molecule BEC as a two mode system, an atomic mode and the another one molecular. This bipartite system is entangled by the interconversion dynamic of atoms and molecules and we can study the quantum phase transition of the system using tools of the quantum information \cite{GSantos10,Angela2}.
 The dynamics of interconversion of atoms and molecules as an open system, with particles losses was studied in \cite{Cui}. We are considering a closed system with the total number of atoms conserved\cite{GSantosCA1}. The dynamic of interconversion in an atom-molecule Bose-Einstein condensate is described by the Hamiltonian 
 
\begin{eqnarray}
\hat{H} & = & U_a\hat{N}_a^2 + U_b\hat{N}_b^2 + U_{ab}\hat{N}_a\hat{N}_b + \mu_a\hat{N}_a +\mu_b\hat{N}_b  -   \Omega(\hat{a}^{\dag}\hat{a}^{\dag}\hat{b} + \hat{b}^{\dag}\hat{a}\hat{a}),
\label{ham}
\end{eqnarray}
\noindent where $\hat{a}^{\dagger}$ and $\hat{a}$ are the creation and annihilation boson operators of an unbound atom while  $\hat{b}^{\dagger}$  and $\hat{b}$ are the creation and annihilation boson operators of a molecule. The number boson operators $\hat{N}_a=\hat{a}^{\dagger}\hat{a}$ and $\hat{N}_b=\hat{b}^{\dagger}\hat{b}$ are the respective number of unbound atoms and number of molecules operators. These bosons operators satisfies the canonical commutation relations
\begin{equation}
 [\hat{a},\hat{a}^\dagger] =  [\hat{b},\hat{b}^\dagger] = \hat{I},  
\end{equation}
\noindent with the condition that the  unbound atom and molecule boson operators commutes
\begin{equation}
 [\hat{a}^\dagger,\hat{b}] = [\hat{a},\hat{b}] = 0,  
\end{equation}
\noindent and
\begin{equation}
 [\hat{N}_a,\hat{a}] =  - \hat{a}, \qquad [\hat{N}_a,\hat{a}^\dagger] =  +\hat{a}^\dagger, 
\end{equation}
\begin{equation}
 [\hat{N}_b,\hat{b}] =  - \hat{b}, \qquad [\hat{N}_b,\hat{b}^\dagger] =  +\hat{b}^\dagger, 
\end{equation}
\noindent where $\hat{I}$ is the identity operator. 

The  parameter  $U_{a}$ is the atom-atom interaction, $U_{ab}$ is the atom-molecule interaction and $U_{b}$ is the molecule-molecule interaction. The strength of these interactions are proportional to the  scattering length and usually is enough to consider only the low energy $s$-wave scattering length.  
The parameter $\mu_a$ is the external potential for the unbound atoms and  $\mu_b$ is the external potential for the molecules. The parameter $\Omega$ is the amplitude for interconversion of atoms and molecules.  
In the limit $U_{a}=U_{ab}=U_{b}=0$, the Hamiltonian (\ref{ham}) has been studied using a
variety of methods \cite{Vardi,Hines,Clare,Shen,Graefe}. However in the experimental context, the $s$-wave scattering interactions play a significant role.  It will be seen below that for the general model (\ref{ham}) the inclusion of these scattering terms has a non-trivial consequence. The $s$-wave scattering length for the atom-atom interaction $U_{a}$ can be determined precisely by photoassociation spectroscopy of one and two photons \cite{Amelink,AnnualReview}. We mention that generally the values for the molecule-molecule $U_b$ and the atom-molecule $U_{ab}$ interactions are unknown \cite{Wynar}, although some estimates exist in the case of $^{85}Rb$ \cite{Cusack}. For an up to dated discussion of theoretical and experimental data see \cite{AnnualReview}. 

The paper is organized as follows. In the section 2, I present the symmetries of the Hamiltonian. In the section 3, I present the interconversion currents algebra and the Casimir operators. In the section 4, I calculate the quantum dynamics of the currents. In the section 5, I summarize the results. 
 
\section{Symmetries}

 The Hamiltonian (\ref{ham}) is invariant under the global $U(1)$  gauge transformation $\hat{a} \rightarrow e^{i\alpha}\hat{a}$, $\hat{b} \rightarrow e^{2i\alpha}\hat{b}$, where $\alpha$ is an arbitrary $c$-number and $\hat{a}^{\dagger}\rightarrow e^{-i\alpha}\hat{a}^{\dagger}$, $\hat{b}^{\dagger}\rightarrow e^{-2i\alpha}\hat{b}^{\dagger}$. The global $U(1)$  gauge invariance is associated with the conservation of the total number of atoms (unbound atoms plus bound atoms) $\hat{N} = \hat{N}_a + 2\hat{N}_b$.  The Hamiltonian (\ref{ham}) is not invariant under the exchange of atoms and molecules because the interconversion dynamics is not symmetric under this exchange.

The Hamiltonian (\ref{ham}) is invariant under the $\mathbb{Z}_2$ mirror transformation  of the unbound atom operators $\hat{a} \rightarrow -\hat{a}, \hat{a}^{\dagger} \rightarrow -\hat{a}^{\dagger}$, but it is not invariant under the $\mathbb{Z}_2$ mirror transformation  of the molecule operators $\hat{b} \rightarrow -\hat{b}, \hat{b}^{\dagger} \rightarrow -\hat{b}^{\dagger}$. The $\mathbb{Z}_2$ symmetry is associated with the parity of the wave function by the relation

\begin{equation}
 \hat{P}_a \; |\Psi\rangle = (-1)^{N_a} |\Psi\rangle,
\end{equation}  
\noindent where $\hat{P}_a$ is the parity operator acting only on $\hat{a}$ and $\hat{a}^{\dagger}$. Because this symmetry we have  $[\hat{H},\hat{P}_a] = 0$. When we consider the parity operator $\hat{P}_b$ acting only  on the  molecule operators $\hat{b}$ and $\hat{b}^{\dagger}$, the parity of the wave function is not well defined and we have $[\hat{H},\hat{P}_b] \neq 0$.

If $N_a$ is even, $N$ is even and the dimension of the state space is $D=N/2 +1$.  For $N_a$  even we can write a general state of the Hamiltonian (\ref{ham}) as

\begin{equation}
|\Psi\rangle = \sum_{n_b = 0}^{\frac{N}{2}} \; C_{N-2n_b,n_b} \frac{(\hat{a}^{\dagger})^{N-2n_b}}{\sqrt{(N-2n_b)!}} \frac{(\hat{b}^{\dagger})^{n_b}}{\sqrt{n_b!}} |0,0\rangle,
\end{equation} 
\noindent and
\begin{equation}
 \hat{P}_a \; |\Psi\rangle = |\Psi\rangle.
\end{equation}
\noindent If $N_a$ is odd, $N$ is odd and the dimension of the state space is $D=(N+1)/2$. For $N_a$ odd we can write a general state of the Hamiltonian (\ref{ham}) as
\begin{equation}
|\Psi\rangle = \sum_{n_b = 0}^{\frac{N - 1}{2}} \; C_{N-2n_b,n_b} \frac{(\hat{a}^{\dagger})^{N-2n_b}}{\sqrt{(N-2n_b)!}} \frac{(\hat{b}^{\dagger})^{n_b}}{\sqrt{n_b!}} |0,0\rangle,
\end{equation} 
\noindent and
\begin{equation}
 \hat{P}_a \; |\Psi\rangle = - |\Psi\rangle.
\end{equation}

The symmetries of the Hamiltonian (\ref{ham}) imply  degeneracy. For the conservancy of $\hat{N}$ we have  that all wave function of the Hamiltonian (\ref{ham}) are degenerated eigenfunctions of $\hat{N}$ with the same eigenvalue $N$.  For the parity operator $\hat{P}_a$ all wave function of the Hamiltonian (\ref{ham}) are even or odd depending if $N$ is even or odd. All wave functions are degenerated eigenfunctions of $\hat{P}_a$ with the same eigenvalue $\lambda = +1$ if $N$ is even  or they are degenerated eigenfunctions of $\hat{P}_a$ with the same eigenvalue $\lambda = -1$ if $N$ is odd.

\section{Interconversion Currents Algebra}


The quantum dynamics of any operator $\hat{O}$ in the Heisenberg picture is determined by the Heisenberg equation of motion
\begin{equation}
 \frac{d\hat{O}}{dt} = \frac{i}{\hbar}[\hat{H},\hat{O}].
 \label{dOdtH}
\end{equation}

The boson operator total number of atoms, $\hat{N} = \hat{N}_a + 2\hat{N}_b$,  is a conserved quantity, $[\hat{H},\hat{N}]=0$, and it is commutable compatible operator (CCO) with the number of unbound atoms and number of molecules boson operators, $[\hat{N},\hat{N}_a]=[\hat{N},\hat{N}_b]= [\hat{N}_a,\hat{N}_b]=0$. The boson operators number of unbound atoms and the number of molecules  don't commute with the Hamiltonian and their time evolution is dictated by the interconversion current operator,

\begin{equation}
\hat{\mathcal{J}} = \frac{1}{4i} (\hat{a}^\dagger \hat{a}^\dagger\hat{b} - \hat{b}^\dagger\hat{a}\hat{a}),
\label{cj}
\end{equation}
\noindent in coherent opposite phases because of the conservancy of $\hat{N}$, with 
\begin{equation}
[\hat{H},\hat{N}_a]= + 8 i\Omega\hat{\mathcal{J}}, ~~~ [\hat{H},\hat{N}_b]= -4i\Omega\hat{\mathcal{J}},
\end{equation}
\noindent and 
\begin{equation}
\frac{d\hat{N}_a}{dt}  =  - 8\frac{\Omega}{\hbar}\hat{\mathcal{J}}, \label{CDN1}
\end{equation}

\begin{equation}
\frac{d\hat{N}_b}{dt}  =  + 4 \frac{\Omega}{\hbar}\hat{\mathcal{J}}. \label{CDN2}
\end{equation}

Integrating the Eqs. (\ref{CDN1}) and (\ref{CDN2}) we get
\begin{eqnarray}
\hat{N}_a(t)  &=&  \hat{N}_a(0) - 8\frac{\Omega}{\hbar}\int_0^t\hat{\mathcal{J}}(\tau) \; d\tau, \label{SCDN1} \\
\hat{N}_b(t) &=& \hat{N}_b(0) + 4\frac{\Omega}{\hbar}\int_0^t\hat{\mathcal{J}}(\tau) \; d\tau. \label{SCDN2}
\end{eqnarray}

   The operator $\hat{\mathcal{J}}$ together with the imbalance current operator $\hat{\mathcal{I}}$,
\begin{equation}
\hat{\mathcal{I}} = \frac{1}{4}(\hat{N}_a - 2\hat{N}_b),
\label{ci}
\end{equation}  
\noindent and the coherent correlation interconversion current operator $\hat{\mathcal{T}}$,
\begin{equation}
\hat{\mathcal{T}} = \frac{1}{4}(\hat{a}^\dagger\hat{a}^\dagger\hat{b} + \hat{b}^\dagger\hat{a}\hat{a}),
\label{ct}
\end{equation}
\noindent  generates the currents algebra
\begin{equation}
 [\hat{\mathcal{T}},\hat{\mathcal{J}}]= +if(\hat{\mathcal{I}},\hat{N}), ~~~ [\hat{\mathcal{T}},\hat{\mathcal{I}}]= -i\hat{\mathcal{J}}, ~~~ [\hat{\mathcal{J}},\hat{\mathcal{I}}]= +i\hat{\mathcal{T}} \nonumber,
\label{calg}
\end{equation}
\noindent with
\begin{equation}
f(\hat{{\cal I}},\hat{N})=\frac{3}{2}\hat{{\cal I}}^{2}+\frac{1}{4}\hat{N}\hat{{\cal I}}-\hat{\mathcal{N}},
\label{calgf}
\end{equation}

\noindent and
\begin{equation}
\hat{\mathcal{N}}=\frac{\hat{N}}{8}\left(\frac{\hat{N}}{4}+1\right).
\end{equation}

\noindent With the identification $\hat{L}_x \equiv  \hat{\mathcal{T}}$, $\hat{L}_y \equiv  \hat{\mathcal{J}}$, and $\hat{L}_z \equiv \hat{\mathcal{I}}$ we can write (\ref{calg}) as the deformed  momentum angular algebra
\begin{equation}
 [\hat{L}_k,\hat{L}_l] = i\varepsilon_{klx}\hat{L}_x + i\varepsilon_{kly}\hat{L}_y +  i\varepsilon_{klz} f(\hat{L}_z,\hat{N}),
\end{equation}
\noindent where $\varepsilon_{klm}$ is the antisymmetric Levi-Civita tensor, with $k,l=x,y,z$ and $\varepsilon_{xyz} = +1$.

 We have two Casimir operators for that current algebra using the deformed momentum angular realization. One of them is the total number of atoms $\hat{N}$, related to the global  $U(1)$ gauge symmetry, $\hat{\mathcal{\mathfrak{C}}}_{1} = \hat{N}$, and the another one is related to the  deformed momentum angular algebra with the broken $O(3)$ symmetry
\begin{eqnarray}
 \hat{\mathcal{\mathfrak{C}}}_{2} &=& \hat{L}_{x}^{2}+\hat{L}_{y}^{2}+\hat{L}_{z}^{3}+\frac{\hat{N}}{4}\hat{L}_{z}^{2}+\frac{1}{2}\left(1-\frac{\hat{N}}{2} -\frac{\hat{N}^2}{8}\right)\hat{L}_{z}. \nonumber\\ \label{CQ2}
\end{eqnarray}
\noindent We can show that $\hat{\mathcal{\mathfrak{C}}}_{2}$ is just a function of $\hat{\mathcal{\mathfrak{C}}}_{1}$
\begin{equation}
\hat{\mathcal{\mathfrak{C}}}_{2} = \frac{\hat{\mathcal{\mathfrak{C}}}^2_{1}}{16}\left(\frac{\hat{\mathcal{\mathfrak{C}}}_{1}}{4} +  1\right).
\end{equation}
\noindent The Casimir operators $\hat{\mathcal{\mathfrak{C}}}_{1}$ and $\hat{\mathcal{\mathfrak{C}}}_{2}$, the boson number of unbound atoms $\hat{N}_a$, the boson number of molecules $\hat{N}_b$,  and the imbalance current operator, $\hat{\mathcal{I}}$, are CCO and so they have the same set of eigenfunctions and can simultaneous have well defined  values

\begin{eqnarray}
\hat{\mathcal{\mathfrak{C}}}_{2} |n_a,n_b\rangle &=& \frac{N^2}{16}\left(\frac{N}{4} +  1\right) |n_a,n_b\rangle, \\
\hat{\mathcal{I}} |n_a,n_b\rangle &=& \frac{1}{4} \left(n_a - 2 n_b\right) |n_a,n_b\rangle.
\end{eqnarray}
We also can use the realization of a deformed $SU(2)$ algebra \cite{Graefe,Korsch,Bonatsos,Watanabe}

\begin{equation}
\hat{\mathcal{L}}_{\pm} = \hat{L}_{x} \pm i\hat{L}_{y}, \qquad
\hat{\mathcal{L}}_{z} = \hat{L}_{z},
\end{equation}
\noindent with the commutation relations
\begin{equation}
[\hat{\mathcal{L}}_{z}, \hat{\mathcal{L}}_{\pm}]  =  \pm \hat{\mathcal{L}}_{+}, \qquad 
[\hat{\mathcal{L}}_{+}, \hat{\mathcal{L}}_{-}]  =   2 f(\hat{\mathcal{L}}_{z},\hat{N}),\nonumber
\end{equation}
\noindent that we can write as

\begin{eqnarray}
[\hat{\mathcal{L}}_{k},\hat{\mathcal{L}}_{l}] &=& \varepsilon_{kl-}\hat{\mathcal{L}}_{+} + \varepsilon_{kl+}\hat{\mathcal{L}}_{-} + 2 \varepsilon_{zkl}f(\hat{\mathcal{L}}_{z},\hat{N}),
\end{eqnarray}
\noindent  with $k,l=z,+,-$ and $\varepsilon_{z+-} = +1$.

This deformed $SU(2)$ algebra has three Casimr operators, $\hat{\mathcal{\mathfrak{C}}}_{1}$,

\begin{equation}
\hat{\mathcal{\mathfrak{C}}}_{3} = \hat{\mathcal{L}}_{+}\hat{\mathcal{L}}_{-} + \hat{\mathcal{L}}_{z}^{3} + \frac{1}{2}\left(\frac{\hat{N}}{2}-3\right)\hat{\mathcal{L}}_{z}^{2} + \frac{1}{2}\left[1-\hat{N}\left(\frac{\hat{N}}{8}+1\right)\right]\hat{\mathcal{L}}_{z}
\label{CQ3}
\end{equation}
\noindent and
\begin{equation}
\hat{\mathcal{\mathfrak{C}}}_{4} = \hat{\mathcal{L}}_{-}\hat{\mathcal{L}}_{+} + \hat{\mathcal{L}}_{z}^{3} + \frac{1}{2}\left(\frac{\hat{N}}{2}+3\right)\hat{\mathcal{L}}_{z}^{2} + \frac{1}{2}\left(1-\frac{\hat{N}^{2}}{8}\right)\hat{\mathcal{L}}_{z}.
\label{CQ4}
\end{equation}

That we can write, using the $\hat{\mathcal{\mathfrak{C}}}_{1}$ Casimir operator, as

\begin{equation}
\hat{\mathcal{\mathfrak{C}}}_{3}=\frac{1}{64}(\hat{\mathcal{\mathfrak{C}}}_{1}^{3}+2\hat{\mathcal{\mathfrak{C}}}_{1}^{2}-8\hat{\mathcal{\mathfrak{C}}}_{1}),
\end{equation}

\begin{equation}
\hat{\mathcal{\mathfrak{C}}}_{4}=\frac{1}{64}(\hat{\mathcal{\mathfrak{C}}}_{1}^{3}+6\hat{\mathcal{\mathfrak{C}}}_{1}^{2}+8\hat{\mathcal{\mathfrak{C}}}_{1}).
\end{equation}

They are related by the equation,

\begin{eqnarray}
\hat{\mathcal{\mathfrak{C}}}_{4}  & = & \hat{\mathcal{\mathfrak{C}}}_{3}+\frac{\hat{\mathcal{\mathfrak{C}}}_{1}}{4}\left(\frac{\hat{\mathcal{\mathfrak{C}}}_{1}}{4}+1\right).
\end{eqnarray}

\begin{figure}
\begin{center}
\begin{tabular}{cc}
$(a)$ & $(b)$ \\
\includegraphics[scale=0.5]{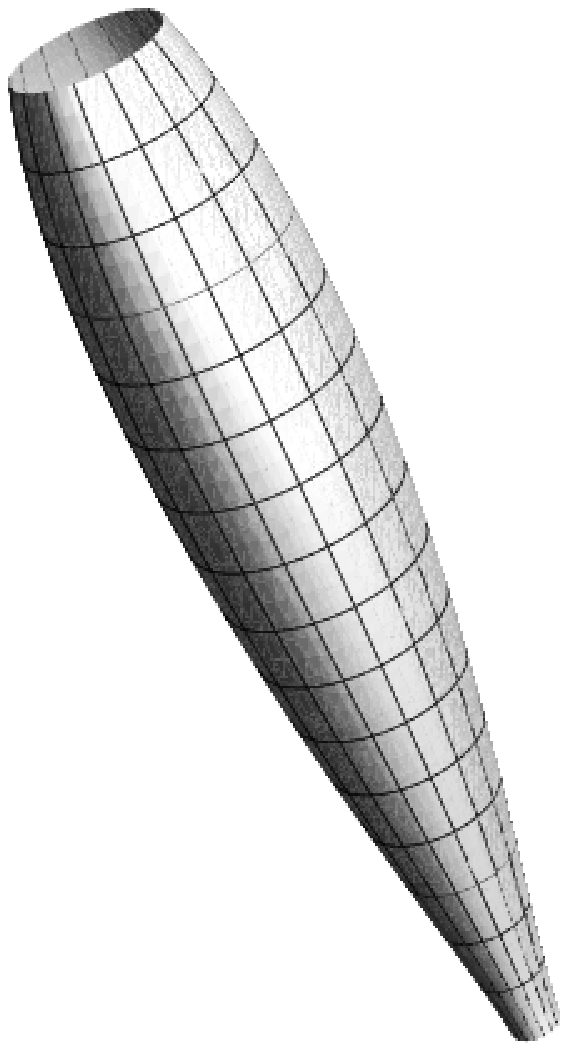} & \includegraphics[scale=0.5]{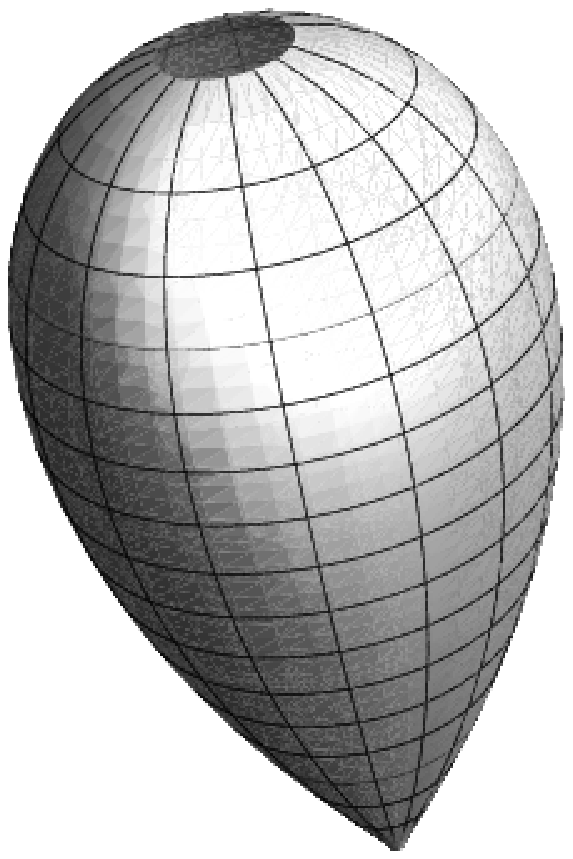} \\
$(c)$ & $(d)$ \\
\includegraphics[scale=0.5]{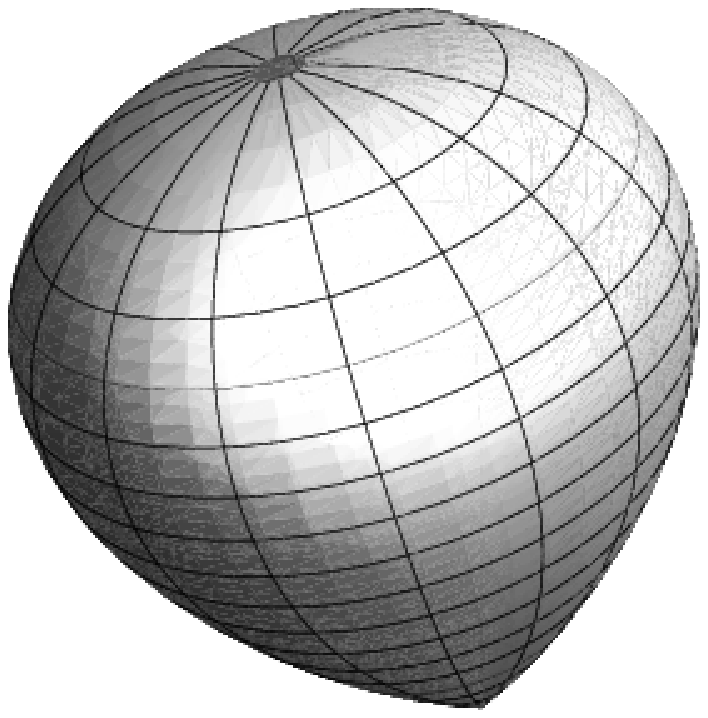} & \includegraphics[scale=0.5]{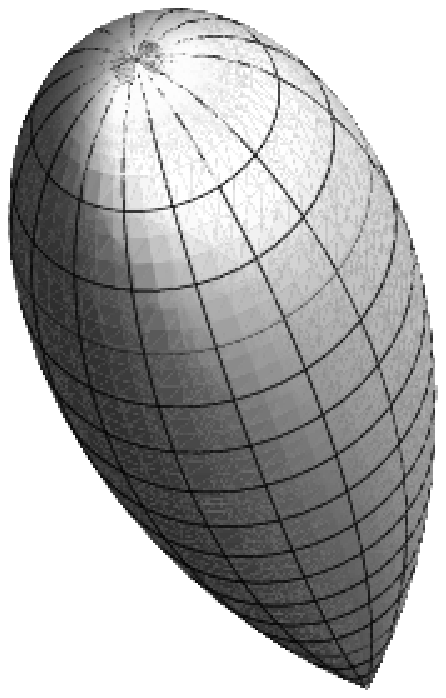} \\
\end{tabular}
\caption{Casimir surfaces for $(a)$ $N = 10$ and $b=10$, $(b)$ $N = 100$ and $b=10$, $(c)$ $N = 1000$ and $b=20$, $(d)$ $N = 15000$ and $b=200$.} 
\label{CSF}
\end{center}
\end{figure}

\begin{figure}
\begin{center}
\begin{tabular}{cc}
$(a)$ & $(b)$ \\
\includegraphics[scale=0.5]{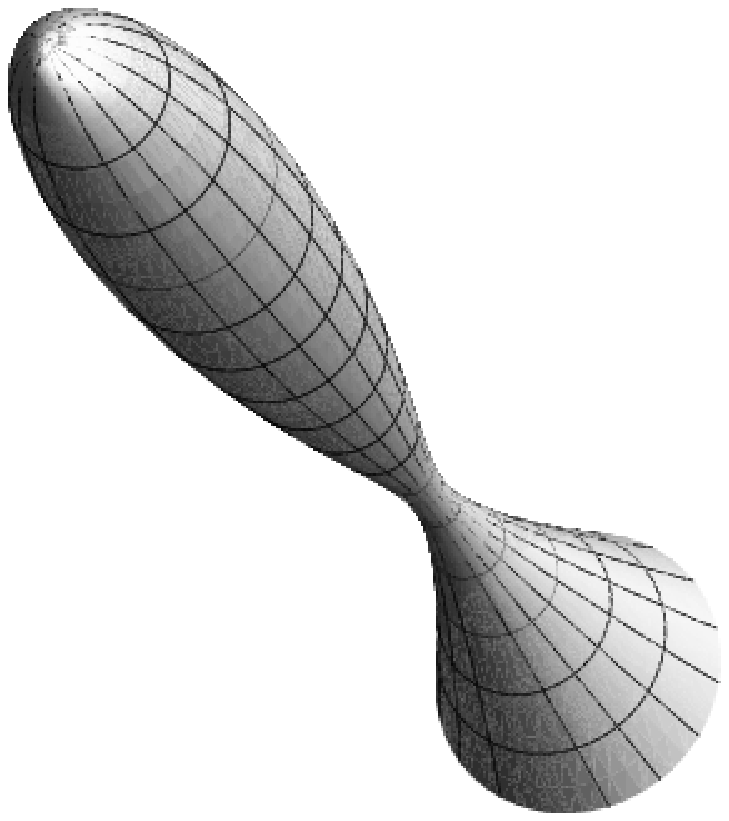} & \includegraphics[scale=0.5]{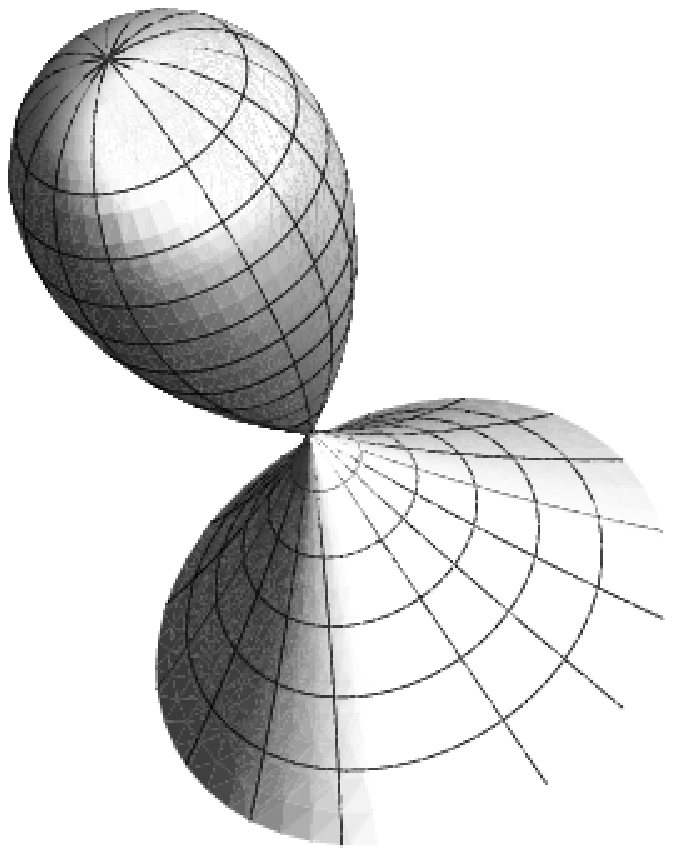} \\
$(c)$ & $(d)$ \\
\includegraphics[scale=0.5]{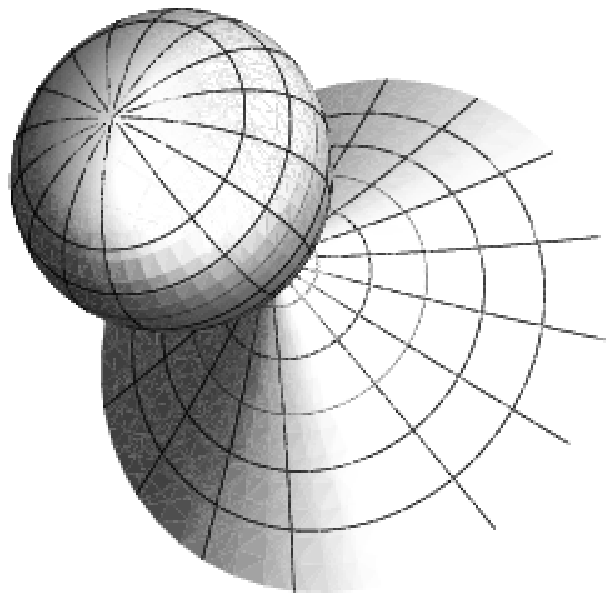} & \includegraphics[scale=0.5]{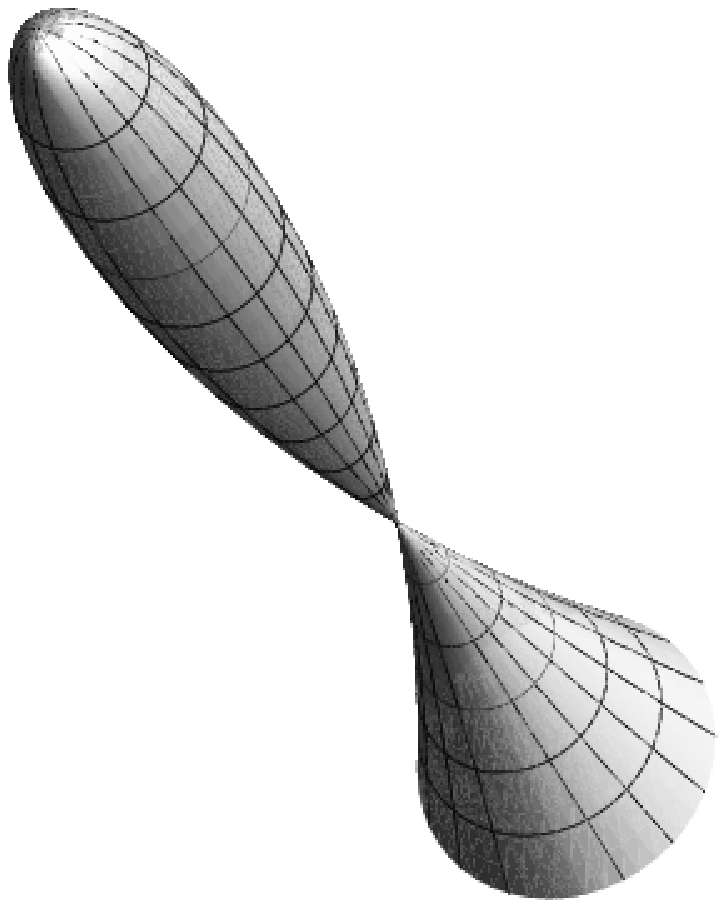} \\
\end{tabular}
\caption{Casimir surfaces for $v\in[-0.55N,0.55N]$ and $(a)$ $N = 10$ and $b=10$, $(b)$ $N = 100$ and $b=10$, $(c)$ $N = 1000$ and $b=20$, $(d)$ $N = 15000$ and $b=100$.} 
\label{CSF2}
\end{center}
\end{figure}

Using the identities

\begin{equation}
\hat{\mathcal{L}}_{+}\hat{\mathcal{L}}_{-} = \hat{\mathcal{L}}_{x}^{2} + \hat{\mathcal{L}}_{y}^{2} + f(\hat{\mathcal{L}}_{z},\hat{N}), \qquad 
\hat{\mathcal{L}}_{-}\hat{\mathcal{L}}_{+} = \hat{\mathcal{L}}_{x}^{2} + \hat{\mathcal{L}}_{y}^{2} - f(\hat{\mathcal{L}}_{z},\hat{N}), \label{ident1}
\end{equation}
\noindent we can write the Cassimir operators (\ref{CQ3}) and (\ref{CQ4}) as

\begin{eqnarray}
\hat{\mathcal{\mathfrak{C}}}_{3} & = & \hat{\mathcal{L}}_{x}^{2}+\hat{\mathcal{L}}_{y}^{2}+\hat{\mathcal{L}}_{z}^{3}+ \frac{\hat{N}}{4}\hat{\mathcal{L}}_{z}^{2}+\frac{1}{2}\left(1-\frac{\hat{N}}{2}-\frac{\hat{N}^{2}}{8}\right)\hat{\mathcal{L}}_{z}-\hat{\mathcal{N}},
\end{eqnarray}

\begin{eqnarray}
\hat{\mathcal{\mathfrak{C}}}_{4} & = & \hat{\mathcal{L}}_{x}^{2}+\hat{\mathcal{L}}_{y}^{2}+\hat{\mathcal{L}}_{z}^{3}+\frac{\hat{N}}{4}\hat{\mathcal{L}}_{z}^{2}+\frac{1}{2}\left(1-\frac{\hat{N}}{2}-\frac{\hat{N}^{2}}{8}\right)\hat{\mathcal{L}}_{z}+\hat{\mathcal{N}}.
\end{eqnarray}
\noindent It is easy to show that $\hat{\mathcal{\mathfrak{C}}}_{4} - \hat{\mathcal{N}} = \hat{\mathcal{\mathfrak{C}}}_{3} + \hat{\mathcal{N}} = \hat{\mathcal{\mathfrak{C}}}_{2}$ and $\hat{\mathcal{\mathfrak{C}}}_{4} - \hat{\mathcal{\mathfrak{C}}}_{3} = 2\hat{\mathcal{N}}$. Therefore, if we take the average value as in  \cite{Graefe,Korsch,Watanabe} we see that the Casimir operators $\hat{\mathcal{\mathfrak{C}}}_{2}$, $\hat{\mathcal{\mathfrak{C}}}_{3} + \hat{\mathcal{N}}$ and $\hat{\mathcal{\mathfrak{C}}}_{4} - \hat{\mathcal{N}}$ describe the same surfaces. Some of these Casimir surfaces has been denoted as Kummer shapes  \cite{Graefe,Korsch,Holm1,Holm2,Kummer1,Kummer2,Kummer3,Kummer4}. We plot these surfaces in the Fig. (\ref{CSF}) using the following parametrization

\begin{eqnarray}
X(u,v) &=& f(v)\; cos\;u, \\
Y(u,v) &=& f(v)\; sin\;u, \\
Z(u,v) &=& b\;v, 
\end{eqnarray}
\noindent where

\begin{equation}
f(v) = \sqrt{C_2 - v^3 - \frac{N}{4}v^2 - \frac{1}{2}\left( 1 - \frac{N}{2} - \frac{N^2}{8} \right)v},
\end{equation}
\noindent with $ C_2 = \frac{N^2}{16}\left(\frac{N}{4} +1\right)$ the eigenvalue of the Casimir operator $\hat{\mathcal{\mathfrak{C}}}_{2}$, $0\leq u \leq 2\pi$ and $-\frac{N}{4} \leq v \leq \frac{N}{4}$. The parameter $b$ changes the scale in the $Z$-axis direction.  The radii in the boundaries of the  $Z$-axis direction are $f(-\frac{N}{4}) = \sqrt{\frac{N}{8}}$ and $f(\frac{N}{4}) = \sqrt{\frac{N}{8}\left(N - 1\right)}$. If we extend the physical limit of $v$ we get the surfaces in Fig. (\ref{CSF2}). 


Using the commutation relations of the currents  (\ref{calg}) it is easy to calculate the anticommutators

\begin{eqnarray}
[\hat{\mathcal{T}},\hat{\mathcal{I}}]_{+} & = & 2 \hat{\mathcal{I}} \hat{\mathcal{T}} - i\hat{\mathcal{J}}, \\ \label{AC1a}
[\hat{\mathcal{T}},\hat{\mathcal{J}}]_{+} & = & 2 \hat{\mathcal{J}} \hat{\mathcal{T}} + i  f(\hat{\mathcal{I}},\hat{N}), \\ \label{AC2a}
[\hat{\mathcal{J}},\hat{\mathcal{I}}]_{+} & = & 2 \hat{\mathcal{I}} \hat{\mathcal{J}} + i\hat{\mathcal{T}}.\label{AC3a}
\end{eqnarray} 
\noindent We will use these anticommutators together with the commutators (\ref{calg}) in the calculus of the currents quantum dynamics.

\section{Interconversion Currents Quantum Dynamics} 

 We can rewrite the Hamiltonian (\ref{ham}) using the currents operators (\ref{cj}), (\ref{ci}) and (\ref{ct}) as
\begin{eqnarray}
\hat{H} &=& \alpha\hat{\mathcal{I}}^2 + \hat{\mathbf{\zeta}}\hat{\mathcal{I}} - 4\Omega\hat{\mathcal{T}} + \rho\hat{\mathcal{\mathfrak{C}}}_{1}^2 + \xi\hat{\mathcal{\mathfrak{C}}}_{1},
\label{ham2} 
\end{eqnarray}
\noindent with the following identification for the parameters
\begin{eqnarray}
\alpha & = & 4U_a + U_b - 2U_{ab}, \nonumber \\
\beta  & = & \left(2U_a - \frac{U_b}{2} \right),\nonumber \\
\gamma & = & 2\mu_a - \mu_b, \nonumber \\
\rho   & = & \left(\frac{U_a}{4} + \frac{U_b}{16} + \frac{U_{ab}}{8} \right), \nonumber \\
\xi    & = & \left( \frac{\mu_a}{2} + \frac{\mu_b}{4} \right), \nonumber
\end{eqnarray}
\noindent and with the definition of the Casimir operator $\hat{\mathbf{\zeta}}  =  \beta\hat{\mathcal{\mathfrak{C}}}_{1} + \gamma$.

The quantum dynamic of the currents are determined by the currents algebra (\ref{calg}), their commutation relations with the Hamiltonian and the parameters. Following \cite{GSantosCA1} we can write the second time derivative of any operator $\hat{O}$ in the Heisenberg picture as
\begin{equation}
 \frac{d^2\hat{O}}{dt^2} = \left(\frac{i}{\hbar}\right)^2 [\hat{H},[\hat{H},\hat{O}]],
 \label{dOdt0}
\end{equation}
\noindent or as
\begin{equation}
 \frac{d^2\hat{O}}{dt^2} = \frac{i}{\hbar}[\hat{H},\frac{d\hat{O}}{dt}].
 \label{dOdt}
\end{equation}

We can see from the Hamiltonian (\ref{ham2}) and Eq. (\ref{dOdt0}) that all Casimir operators are conserved quantities, $[\hat{H},\hat{\mathcal{\mathfrak{C}}}_{1}]=0$. Because the operator $\hat{N}$ is  a Cassimir operator and a conserved quantity all wave function of the Hamiltonians (\ref{ham}) or (\ref{ham2}) are eigenfunctions of that operator with degenerated eigenvalue $N$ \cite{GSantosBVS-PLB}. As the another Casimir operators are functions of $\hat{N}$  we have also that all wave function of the Hamiltonians are degenerated eigenfunctions of these operators with the respective eigenvalues as functions of $N$.  Therefore, we will consider the Casimir operators $\hat{\mathbf{\zeta}}$ and $\hat{N}$ as the respective $c$-numbers $N$ and $\zeta =  \beta N + \gamma$ in the calculus of the currents quantum dynamics. So we also will write $\hat{\mathcal{N}}$ and $f(\hat{{\cal I}},\hat{N})$ as the respective functions of $N$,

\begin{equation}
 \mathcal{N} = \frac{N}{8}\left(\frac{N}{4}+1\right), \qquad f(\hat{{\cal I}},N) = \frac{3}{2}\hat{{\cal I}}^{2}+\frac{N}{4}\hat{{\cal I}}-\mathcal{N}.
\label{c-n} 
\end{equation}

Using the Eq. (\ref{ham2}) and (\ref{dOdt0}) or (\ref{dOdt}) we found the following equations for the quantum dynamics of the three currents

\begin{eqnarray}
\frac{d^{2}\hat{\mathcal{I}}}{dt^{2}}+i\frac{\alpha}{\hbar}\frac{d\hat{\mathcal{I}}}{dt}+24\frac{\Omega^{2}}{\hbar^{2}}\hat{{\cal I}}^{2}+4\frac{\Omega^{2}}{\hbar^{2}}N\hat{{\cal I}} & = & - 8\frac{\Omega\alpha}{\hbar^{2}}\hat{\mathcal{I}}\hat{\mathcal{T}} - 4\frac{\Omega\zeta}{\hbar^{2}}\hat{\mathcal{T}} + 16\frac{\Omega^{2}}{\hbar^{2}}\mathcal{N}, 
\label{eq1:wideeq}
\end{eqnarray}

\begin{eqnarray}
\frac{d^{2}\hat{\mathcal{J}}}{dt^{2}}+\frac{1}{\hbar^{2}}\left[\alpha^{2}+4\Omega^{2}N+\zeta^{2}\right]\hat{\mathcal{J}} & = & -4\frac{\alpha^{2}}{\hbar^{2}}\hat{\mathcal{I}}^{2}\hat{\mathcal{J}} - 4i\frac{\alpha^{2}}{\hbar^{2}}\hat{\mathcal{I}}\hat{\mathcal{T}} - 8\frac{\Omega\alpha}{\hbar^{2}}\hat{\mathcal{J}}\hat{\mathcal{T}} \nonumber\\
 & - & \frac{4}{\hbar^{2}}\left(\alpha\zeta+12\Omega^{2}\right)\hat{{\cal I}}\hat{\mathcal{J}}-\frac{2i}{\hbar^{2}}\left(\alpha\zeta+12\Omega^{2}\right)\hat{\mathcal{T}} \nonumber\\
 & - & 6i\frac{\Omega\alpha}{\hbar^{2}}\hat{{\cal I}}^{2} - i\frac{\Omega\alpha}{\hbar^{2}}N\hat{{\cal I}} + 4i\frac{\Omega\alpha}{\hbar^{2}}\mathcal{N},
\label{eq2:wideeq}
\end{eqnarray}

\begin{eqnarray}
\frac{d^{2}\hat{\mathcal{T}}}{dt^{2}}+\frac{1}{\hbar^{2}}\left[\alpha^{2}+\zeta^{2}\right]\hat{\mathcal{T}} & = & -4\frac{\alpha^{2}}{\hbar^{2}}\hat{\mathcal{I}}^{2}\hat{\mathcal{T}}+4i\frac{\alpha^{2}}{\hbar^{2}}\hat{\mathcal{I}}\hat{\mathcal{J}}-4\frac{\alpha\zeta}{\hbar^{2}}\hat{\mathcal{I}}\hat{\mathcal{T}}+2i\frac{\alpha\zeta}{\hbar^{2}}\hat{\mathcal{J}} \nonumber\\
 & + & 8\frac{\Omega\alpha}{\hbar^{2}}\hat{\mathcal{J}}^{2} - 8\frac{\Omega\alpha}{\hbar^{2}}\hat{\mathcal{I}}f(\hat{{\cal I}},N) - 4\frac{\Omega\zeta}{\hbar^{2}}f(\hat{{\cal I}},N).
\label{eq3:wideeq}
\end{eqnarray}

 
\noindent Different choices of the parameters   gives us different dynamics for the currents. The parameters $\rho$ and $\xi$ don't change the dynamics of the currents because they are coupling the Cassimir operator $\hat{\mathcal{\mathfrak{C}}}_{1}$. For $\alpha = \zeta = 0$, the current $\hat{\mathcal{T}}$ is a conserved  quantity, $[\hat{\mathcal{H}},\hat{\mathcal{T}}]=0$, but this don't means that we don't have interconversion dynamics. We can see from Eqs. (\ref{SCDN1}) and (\ref{SCDN2}) that the quantum dynamic of $\hat{N}_a$, $\hat{N}_b$, and $\hat{\mathcal{I}}$ only depend of the interconversion current $\hat{\mathcal{J}}$ and the interconversion parameter $\Omega$. We  only  have dynamics if $\Omega \neq 0$.  We get the following relations between the interactions and external potentials parameters

\begin{equation}
 U_{ab} = 2U_a + \frac{U_b}{2}, \qquad \frac{\mu_b - 2\mu_a}{N} = 2U_a - \frac{U_b}{2}.
\label{cpar1}
\end{equation}
   
\noindent    The simplest choice is the non interaction limit with all $U$'s equal to zero and with all external potentials $\mu$'s equal to zero, but the equations for the currents quantum dynamics are the same since that the parameters obey the relations (\ref{cpar1}).    For $\alpha = 0$ and $\zeta \neq 0$, we have yet the external potentials $\mu$'s and the atom-atom $U_a$,   molecule-molecule $U_b$ and atom-molecule $U_{ab}$ interactions, but the interactions are constrained by the first equation in (\ref{cpar1}). We can also choose $2\mu_a = \mu_b$ or $U_a = U_b/4$ and we will get $\zeta \neq 0$.

\section{Summary}

I have discussed the symmetries of the model and calculated the Casimir operators for the $O(3)$ and $SU(2)$ deformed algebras and showed that the surfaces corresponding to these Casimir operators are the same. I have showed that an interconversion currents algebra appears when I calculate the quantum dynamics of the unbound atoms $\hat{N}_a$ and molecules $\hat{N}_b$ boson number operators. I have used the Heisenberg equation of motion to write the second time derivative of the currents operators. Then I calculated the quantum dynamics of these currents and showed that different dynamics appear when I consider different choices of the parameters of the Hamiltonian. 

\section*{Acknowledgments}

The author acknowledge CAPES/FAPERJ (Coordena\c{c}\~ao de Aperfei\c{c}oamento de Pessoal de N\'{\i}vel Superior/Funda\c{c}\~ao de Amparo \`a Pesquisa do Estado do Rio de Janeiro) for the financial support.



\begin{thebibliography}{10}


\bibitem{ak} J. R. Anglin  and  W. Ketterle, \textit{Nature} \textbf{416} (2002) 211.

\bibitem{anderson} M. H. Anderson, J. R. Ensher, M. R. Mathews, C. E. Wieman  and E. A. Cornell, \textit{Science}  {\bf 269} (1995) 198.

\bibitem{wwcch} J. Williams, R. Walser,  J. Cooper, E. A. Cornell  and M. Holland, \textit{Phys. Rev. } \textbf{A 61} (2000) 0336123.


\bibitem{bose} S. N. Bose, \textit{Z. Phys.}  \textbf{26} (1924) 178.

\bibitem{eins} A. Einstein, \textit{Phys. Math. K1}  \textbf{22} (1924) 261.

\bibitem{Hulet} C. A. Sackett, C. C. Bradley,  M. Welling and R. G. Hulet, \textit{Braz.  Jour. Phys.} \textbf{27} no. 2 (1997) 154.  

\bibitem{Dalfovo} F. Dalfovo, S. Giorgini, L. P. Pitaevskii  and S. Stringari, \textit{Rev. Mod. Phys.}  \textbf{71} (1999) 463.

\bibitem{l01} A. J. Leggett, \textit{Rev. Mod. Phys.}  \textbf{73} (2001) 307.

\bibitem{Bagnato} P. W. Courteille, V. S. Bagnato and  V. I. Yukalov, \textit{Laser Phys.} \textbf{11} (2001) 659.

\bibitem{cw} E. A. Cornell and C. E. Wieman, \textit{Rev. Mod. Phys.} \textbf{74} (2002) 875.

\bibitem{Piza} A. F. R. T. Piza, \textit{Braz. Jour. Phys.} \textbf{34} n. 3B (2004) 1102.

\bibitem{Bloch} I. Bloch, J. Dalibard and W. Zwerger, \textit{Rev. Mod. Phys.}  \textbf{80} (2008) 875.

\bibitem{Carusotto} I. Carusotto and  C. Ciuti, \textit{Rev. Mod. Phys.}  \textbf{85} (2013) 299.

\bibitem{Donley} E. A. Donley, N. R. Claussen, S. T. Thompson, and C. E. Wieman, \textit{ Nature} \textbf{417} (2002) 529.

\bibitem{Zoller}  P. Zoller, \textit{Nature} \textbf{417} (2002) 493.

\bibitem{Souza} N. Vanhaecke, W. de Souza Melo, B. L. Tolra, D. Comparat, and P. Pillet, \textit{Phys. Rev. Lett.} \textbf{89} (2002) 063001.

\bibitem{Wynar} R. Wynar, R.S. Freeland, D.J. Han, C. Ryu, D.J. Heinzen, \textit{Science} \textbf{287}   (2000) 1016.

\bibitem{Richter}  F. Richter, D. Becker, C. B\'eny, T. A. Schulze, S. Ospelkaus  and  T. J. Osborne,  \textit{New J. Phys.} \textbf{17} (2015) 055005.

\bibitem{Hutson} J. M. Hutson, \textit{Science} \textbf{327} (2010) 788.

\bibitem{Regal} C. A. Regal, C. Ticknor, J. L. Bohn  and D. S. Jin, \textit{Nature} \textbf{424} (2003) 47.

\bibitem{Ospelkaus} S. Ospelkaus, K.-K. Ni,  D. Wang,  M. H. G. de Miranda, B. Neyenhuis, G. Qu\'em\'ener, P. S. Julienne, J. L. Bohn, D. S. Jin and J. Ye, \textit{Science} \textbf{327} (2010) 1853.

\bibitem{Heinzen} D. J. Heinzen, R. Wynar, P. D. Drummond  and K. V. Kheruntsyan, \textit{Phys. Rev. Lett.} \textbf{84} (2000) 5029.

\bibitem{Goral} K. G\'oral, M. Gajda,  and K. Rza$\dot{z}$ewski, \textit{Phys. Rev. Lett.} \textbf{86} (2001) 1397.

\bibitem{Carr} L. D. Carr, D. DeMille, R. V. Krems and  J. Ye, \textit{New J. Phys.}  \textbf{11} (2009) 055049.

\bibitem{Kevin} K. M. Jones, E. Tiesinga, P. D. Lett and P. S. Julienne,\textit{Rev. Mod. Phys.} \textbf{78} (2006) 483.

\bibitem{Makrides} C. Makrides, J. Hazra, G. B. Pradhan, A. Petrov, B. K. Kendrick,  Gonz\'alez-Lezana, T., Balakrishnan, N. and  Kotochigova, S., \textit{Phys. Rev. A} \textbf{91} (2015)  012708.

\bibitem{Dulieu}  O. Dulieu and  C. Gabbanini, \textit{Rep. Prog. Phys.} \textbf{72} (2009) 086401.

\bibitem{Pillet}  P. Pillet, N. Vanhaecke,  C. Lisdat, D. Comparat, O. Dulieu, A. Crubellier and  F. Masnou-Seeuws, \textit{Physica Scripta} \textbf{T105} (2003) 7.

\bibitem{Amelink} A. Amelink  and P. van der Straten,  \textit{Physica Scripta} \textbf{68} (2003) C82.

\bibitem{Pillet2} P. Pillet, \textit{Physica Scripta} \textbf{68} (2003) C48. 

\bibitem{Durr}  S. D\"urr, T. Volz, A. Marte and G. Rempe, \textit{Phys. Rev. Lett.}  \textbf{92} (2004) 020406.

\bibitem{GSantos09} A. P. Tonel, C. C. N. Kuhn, G. Santos, A. Foerster, I. Roditi and Z. V. T. Santos, \textit{Phys. Rev. } \textbf{A 79} (2009) 013624.


\bibitem{GSantos06a} G. Santos, A. Tonel, A. Foerster and J. Links, \textit{Phys. Rev. } \textbf{A 73} (2006) 023609.

\bibitem{GSantos10} G. Santos, A. Foerster, J. Links, E. Mattei  and S. R. Dahmen, \textit{Phys. Rev. } \textbf{A 81} (2010)  063621.


\bibitem{GSantos11a} G. Santos, A. Foerster,  I. Roditi, Z. V. T. Santos and A. P. Tonel, \textit{J. Phys. A: Math. Theor.} \textbf{41} (2008) 295003 (9pp).
 
\bibitem{GSantos11b} G. Santos, \textit{J. Phys. A: Math. Theor.} \textbf{44} (2011) 345003.

\bibitem{jlletter} J. Links  and H.-Q. Zhou, \textit{Lett. Math. Phys.} \textbf{60} (2002) 275.

\bibitem{jlreview} J. Links,  H.-Q. Zhou, R. H. McKenzie and M. D. Gould, \textit{J. Phys. A: Math. Gen.} \textbf{36} (2003) R63. 

\bibitem{jinter} H.-Q. Zhou, J. Links  and R. H. Mckenzie, \textit{Int. J. Mod. Phys. B} \textbf{17} (2003) 5819.

\bibitem{jlsigma}  J. Links and K. E. Hibberd, \textit{SIGMA} \textbf{2} (2006) 095 (8pp).

\bibitem{Angela} A. Foerster, J. Links   and  H.-Q. Zhou,
in: {\it Classical and quantum nonlinear integrable systems: theory and applications}, edited
by A. Kundu (Institute of Physics Publishing, Bristol and Philadelphia, 2003)
pp 208--233. 

\bibitem{ATonel} A. P. Tonel and L. H. Ymai, \textit{J. Phys. A: Math. Theor.} \textbf{46} (2013) 125202 (14pp).

\bibitem{GSantos06b}  J. Links, A. Foerster, A. P. Tonel and G. Santos, \textit{Ann. Henri Poincar\'e}  \textbf{7}  (2006) 1591.


\bibitem{Angela2}  D. Rubeni,  A. Foerster,  E. Mattei and I. Roditi, \textit{Nuc. Phys. } \textbf{B 856} (2012) 698.

\bibitem{GSantosBVS-PLB}  G. Santos, C. Ahn, A. Foerster and I. Roditi, \textit{Phys. Lett.} \textbf{B 746} (2015) 186.

\bibitem{jdensity}  J. Links and I. Marquette, \textit{J. Phys. A: Math. Theor.} \textbf{48} (2015) 045204 (15pp).

\bibitem{jbroots} Y. Shen and J. Links, \textit{J. Phys.: Conf. Ser.} \textbf{597} (2015) 12068.

\bibitem{GSantos13} G. Santos, A. Foerster and I. Roditi, \textit{J. Phys. A: Math. Theor.} \textbf{46} (2013) 265206 (12pp).

\bibitem{GSantosCA1} G. N. Santos Filho, \textit{Current algebra for the two-site Bose-Hubbard model},  arXiv:1505.06793 [cond-mat.quant-gas].

\bibitem{GSantosCA2} G. N. Santos Filho, \textit{Current algebra for a generalized two-site Bose-Hubbard model},  arXiv:1511.05026 [cond-mat.quant-gas].


\bibitem{Motohashi1}  A. Motohashi and T. Nikuni, \textit{J. Low Temp. Phys.} \textbf{158}
(2010) 72.

\bibitem{Motohashi2}  A. Motohashi, \textit{Phys. Rev. A} \textbf{84} (2011)  063631.

\bibitem{Motohashi3}  A. Motohashi and T. Nikuni, \textit{Phys. Rev. A} \textbf{82} (2010) 033631.

\bibitem{Motohashi4} A. Motohashi  and T. Nikuni, \textit{J. Phys.: Conf. Ser.} \textbf{ 150} (2009) 032067.

\bibitem{Dukelsky}  A. Rela\~no,  J. Dukelsky,  P. P\'erez-Fern\'andez   and J. M. Arias, \textit{Phys. Rev. E} \textbf{90} (2014) 042139. 

\bibitem{Cui} B. Cui, L. C. Wang and X. X. Yi, \textit{Phys. Rev. A} \textbf{85} (2012) 013618.

\bibitem{Vardi} A. Vardi, V. A. Yurovsky, and J. R. Anglin, \textit{Phys. Rev. A} \textbf{64} 
(2001) 063611.

\bibitem{Hines} A. Hines, R. H. McKenzie and G. J. Milburn, \textit{Phys. Rev. A} \textbf{
67} (2003) 013609.

\bibitem{Clare}   C. Dunning, K. E. Hibberd and J. Links, \textit{J. Stat. Mech.} (2006) P11005.

\bibitem{Shen} H. Z. Shen, X-M Xiu and X. X. Yi, \textit{Phys. Rev. A}
\textbf{87} (2013) 063613.

\bibitem{Graefe} E.-M. Graefe, M.  Graney  and A. Rush,  
\textit{Phys. Rev. A} \textbf{92} (2015) 012121.


\bibitem{Korsch} Eva-Maria Graefe, Hans J\"urgen Korsch and Alexander Rush, \textit{Phys.  Rev.} \textbf{A 93} (2016) 042102.

\bibitem{Holm1} D. D. Holm, \textit{Geometric Mechanics Part I: Dynamics and
Symmetry}, Imperial College Press, London, 2011.

\bibitem{Holm2} D. D. Holm and C. Vizman, \textit{J. Geom. Mech.} \textbf{4} (2012) 297.

\bibitem{Kummer1} M. Kummer, \textit{Comm. Math. Phys.} \textbf{48} (1976) 53.

\bibitem{Kummer2} M. Kummer, \textit{Comm. Math. Phys.} \textbf{58} (1978) 85.

\bibitem{Kummer3} M. Kummer, in \textit{Local and Global Methods in Nonlinear
Dynamics, Lecture notes in Physics, Vol. 252}, edited by A. V. S\'aenz, page 19. Springer, New York, 1986.

\bibitem{Kummer4} M. Kummer, \textit{J. Diff. Eq} \textbf{83} (1990) 220.

\bibitem{Bonatsos} D. Bonatsos, P. Kolokotronis,  C. Daskaloyannis,  A. Ludu  
and C. Quesne,  \textit{Czech. J. Phys.} \textbf{46} (1996) 1189.

\bibitem{Watanabe} A. P. Itin and S. Watanabe, \textit{Phys. Rev. E} \textbf{76} (2007) 026218.

\bibitem{AnnualReview} \textit{Annual Review of Cold Atoms and Molecules}, v. 1, 2 and 3, edited by Kirk W. Madison \textit{et al} (World Scientific, 2013-2015). 

\bibitem{Cusack} B. J. Cusack, T. J. Alexander, E. A. Ostrovskaya and Y. S. Kivshar, \textit{Phys. Rev. A} \textbf{65} (2002) 013609. 



\end{thebibliography}
\end{document}